\documentclass[singlespacing]{elsart}
\usepackage{amssymb}
\usepackage{graphicx}
\usepackage{amsmath}

\begin{document}
\begin{frontmatter}
\title{Translational and rotational mode coupling in disordered
ferroelectric KTa$_{1-x}$Nb$_{x}$O$_{3}$ studied by Raman spectroscopy}
\author{Oleksiy Svitelskiy\corauthref{cor}\thanksref{leave}},
\corauth[cor]{Corresponding author. Fax: (610)7585730}
\ead{ovs2@lehigh.edu}
\thanks[leave]{On leave from the Institute of Semiconductor Physics, Kyiv 03650, Ukraine}
\author{Jean Toulouse}
\address{Physics Department of Lehigh University, Bethlehem, PA 18015, USA}

\begin{abstract}
The coupling of translational modes to the reorientational motion is an
essential property of systems with internal orientational degrees of
freedom. Due to their high complexity most of those systems (molecular
crystals, glasses...) present a major puzzle for scientists. In this paper
we analyze the Raman scattering of a relatively simple ferroelectric system,
KTa$_{1-x}$Nb$_{x}$O$_{3}$, which may serve as a model for more complicated
cases. We show the presence of a strong coupling between translational
and reorientational motion in the crystal. Our data suggest that this
coupling is the main reason for the depolarized component of the second
order Raman spectra and that it is also responsible for the
frequency decrease (softening) of the transverse acoustic mode down to the third of
three transitions, below which reorientational motion is no longer allowed.
\end{abstract}
\begin{keyword}
D. Ferroelectricity \sep D. Lattice dynamics \sep D. Order-disorder effects \sep C. Raman spectroscopy 
\sep Relaxor ferroelectrics
\PACS 77.84.Dy \sep 77.80.-e \sep 63.20.Mt \sep 78.30.-j
\end{keyword}
\end{frontmatter}
\section{Introduction}

Much of the recent ferroelectric research has been focused on the new and
industrially promising lead-based relaxor materials. Unfortunately, from a
basic point of view, these materials are highly complex, presenting both
chemical and structural local order, such as in PbMg$_{1/3}$Nb$_{2/3}$O$_{3}$ \cite{SinyRev} or PbZn$_{1/3}$Nb$_{2/3}$O$_{3}$ \cite{lebon} . Alternatively, there exist lead-free mixed perovskites such as KTa$_{1-x}$Nb$_{x}$O$_{3}$ (KTN) or K$_{1-x}$Li$_{x}$TaO$_{3}$ (KLT), which
exhibit similar properties but are much less complicated than the lead
relaxors \cite{grace,radha} and therefore better suited for an investigation
into the fundamental origin and mechanism of the relaxor behavior\footnote{Besides alkali perovskites, there are some other mixed crystals, like SrTiO$_{3}$ doped with Bi, or BaTiO$_{3}$ doped with Sn, Ce, Zr, etc., that can be used as model relaxors.}. In KTN and KLT, the single most important feature that can lead to the relaxor behavior is the off-centering of Nb \cite{15} or Li \cite{gr11,gr12} and the
resulting formation of polar nanoregions. One of us in his previous work
have shown that these regions and their capability of reorientational
dynamics, are responsible for the first-order Raman scattering \cite{16,17,18,19,20}, 
the unusual softening of the elastic constants, the existence,
even in the paraelectric phase, of the dielectric polarization hysteresis
loops and their frequency dispersion, the pretransitional diffuse neutron
scattering and, finally, for the original coupling between polarization and
strain in the relaxor materials. The mechanism of this coupling was the
best evidenced in our recent studies of remarkable dielectric resonances
observed in KTN and KLT \cite{radha,rada-res,toul-res}, as well as in PbMg$_{1/3}$Nb$_{2/3}$O$_{3}$ \cite{radha,rada-res}. In
the present study, we want to take a new look at the Raman spectra of these
systems, in the light of this polarization-strain coupling or, more
precisely, of the interaction between acoustic phonons and the orientational
motion. This interaction represents a kind of a
rotational-translational (R-T) coupling that is common for \textit{any}
system with orientational disorder (including molecular crystals and
glasses).

\subsection{KTN as a ferroelectric with orientational disorder}

The host system, KTaO$_{3}$, is a highly polarizable paraelectric material,
that has a cubic perovskite structure with Ta ions occupying the centers of
the cells. It does not undergo any phase transition but a smallest admixture
of either Li or Nb causes one or more. In particular, Nb ions replacing Ta
go off-center in a $<$111$>$ direction, endowing the cell with a permanent
electric dipole moment \cite{15}. In the high-temperature phase, these
moments are randomly distributed and reorient rapidly among eight equivalent 
$<$111$>$ directions so that, on average, the cubic symmetry of the lattice
is preserved. With lowering of the temperature, the growing role of the
dipole-dipole interactions leads to the formation of precursor clusters
(polar nanoregions) characterized by permanent giant electric dipole moments
and local distortions from the cubic symmetry \cite{16,17,18,19,20}. If
the concentration of Nb is sufficiently high ($x\gtrsim 5\%$) then, when the
temperature reaches certain critical values $T_{c1},T_{c2},T_{c3}$, the
development of these clusters causes the crystal to undergo a
cubic-tetragonal-orthorhombic-rhombohedral (C-T-O-R) sequence of phase
transitions. If $0.8\lesssim x\lesssim 5\%$, the crystal goes from the C
phase directly to a R phase. If $x\lesssim 0.8\%$, no transition but a
freezing occurs \cite{29}.

In the ideal cubic structure, all zero wavevector phonons are of odd parity,
so first order Raman scattering is forbidden \cite{21} and the spectrum
consists only of second-order scattering peaks. However, the structural
distortions that accompany the formation of the polar nanoregions break the
local cubic symmetry and allow for the appearance and growth (from, approximately $T_{c1}+20$~K) \cite{16,17,18,19,20} of first
order scattering from the transverse polar optic modes: hard TO$_{2}$, TO$%
_{4}$ and soft TO$_{1}$. At $T_{c1}$, the non-polar hard optical mode TO$%
_{3} $ also appears, marking the first structural transition. Throughout
this process, the frequencies of the hard modes remain unchanged but the
frequency of the soft mode continuously decreases and reaches a minimum in a
vicinity of the phase transition \cite{22,22a,22b,22c,22d,22e}.

In KTN, the ferroelectric soft mode, which is a general property of crystals
undergoing a displacive phase transition, coexists with a central peak,
which is a common feature of order-disorder phase transitions. Among the
several models proposed to explain the central peak \cite{lee,lyons,23,23a,23b}, 
the most promising one is based on the orientational disorder of
Nb-ions moving among eight equivalent sites, offered by Sokoloff \cite{23,23a,23b}. While the C phase is characterized by equal probabilities of
occupation of all eight sites, phases of lower symmetry restrict the number
of equivalent sites that are accessible for a given Nb ion to four
neighboring sites in the T phase, to two sites in the O phase and, finally,
to a single site in the R phase. The intersite orientational motion is,
therefore, a property of the C, T and O phases, albeit increasingly more
restricted, but vanishes in the rhombohedral phase. It has a tunneling
rather than a hopping character, and gives rise to the Raman central peak.
Unfortunately, Sokoloff restricted himself to study of the CP behavior only
in a vicinity of the C-T phase transition, so his work is far of being
complete.

In the present paper, we show the results of our light scattering
study of the interaction of this reorientational motion and acoustic modes.
This interaction has so far not been discussed in studies of relaxor
ferroelectrics. But, in our view, it lies at the heart of the unusual
polarization-strain coupling in relaxors and is central for understanding
the relaxor mechanism. The idea of this work originates from earlier studies
of softening of elastic constants $c_{11}$ and $c_{44}$ in KTN on approach
to the transition, which could not be explained simply by TO-TA mode
coupling theory that worked successfully for pure ferroelectrics \cite{axe,prater}. Especially puzzling was the behavior of the constant $c_{11}$,
since it corresponds to a longitudinal distortion and, therefore, not
expected to couple to a transverse optic mode. To resolve this
contradiction, one of us proposed \cite{27} that the softening of these
constants was caused by the interaction of acoustic phonons with reorienting
polar clusters. Their capability to the reorientational dynamics was
evidenced by study of the temperature and frequency dependencies of the
dielectric hysteresis loops that appear together with the polar nanoregions.
The formation of these regions is also accompanied by the appearance and
growth of dielectric resonances that are the direct consequences of a
coupling between polarization and strain \cite{radha,rada-res,toul-res}. In the same temperature range, the
dielectric constant, while still increasing, falls below the values
predicted by Curie-Weiss law. Thus, even though the dielectric constant
indicates a reduction in softening of the system upon approaching the
transition, the polarization-strain coupling increases. Moreover, these
resonances continue to be observed below the first transition, i.e. when the
soft mode frequency increases. This suggests the continuation of the
coupling down to the third (O-R) transition, when rotations stop. Some of
the consequencies of coexistence of the ferroelectric soft mode with the
reorienting polar regions for the dielectric constant of mixed
ferroelectrics have received quantitative evaluation in works 
of Prosandeev et al. \cite{Prosandeev,Prosandeev1,Prosandeev2}
 It is reasonable to expect that the reorientational motion of these
polar nanoregions in the result of a strong polarization-strain coupling
will interact and modify the translational motion in crystal and will be
observable through the Raman spectra. This motion might also be responsible
for disagreement at the zone boundary between measured TA phonon energy and its
theoretically predicted value (see Fig.7 in ref. \cite{22d}).

\subsection{Comparison with cyanides and Michel's theory} 

In developing the present interpretation, we have also drawn from the
understanding that has been gained through investigations of another type of
materials with internal orientational degrees of freedom, namely,
alkali-halide-cyanide compounds (like (KCN)$_{x}$(KBr)$_{1-x}$ or (KCN)$_{x}$%
(KCl)$_{1-x}$...). In their high-temperature phase,
these compounds have an f.c.c. cubic structure
(rocksalt) in which the CN$^{-}$ molecules occupy halogen sites and rapidly
reorient among equivalent $<$111$>$ directions \cite{1}. For sufficiently
high CN$^{-}$ concentrations, lowering temperature causes a transition
to either an orthorhombic or a monoclinic phase (depending on \textit{x}),
with the CN$^{-}$ molecules oriented along one of the $<$110$>$ directions
of the cubic phase \cite{2,3}. Most important for our purpose, this
transition is accompanied by the softening of the transverse acoustic (TA)
mode and, correspondingly, of the $c_{44}$ elastic constant. For low CN$^{-}$
concentrations, (\textit{x}$<$0.56 for (KCN)$_{x}$(KBr)$_{1-x}$ or \textit{x}%
$<$0.8 in (KCN)$_{x}$(KCl)$_{1-x}$), the crystal does not undergo any
transition but, upon lowering temperature, exhibits a freezing of the
dipole orientations \cite{1,4,5}. In both concentration regimes, the TA mode
frequency reaches a minimum at the temperature below which the rotations
cease, i.e. at the phase transition or freezing temperature. Yet, it is
essential to note that these alkali-halide-cyanide compounds do not exhibit
a soft optic mode as does KTN.

The alkali-halide cyanide system is similar to KTN in that, in the
high-temperature phase, both types of compounds possess an average cubic
symmetry characterized by orientational dynamical disorder. As the
temperature is lowered and as the crystalline structure changes, this
dynamic disorder becomes increasingly more static. Numerous studies have
shown that the cyanide system \cite{6} and its Raman spectra \cite{7,8} can be
successfully described by the extended Devonshire model of a rotator in a
cubic field \cite{9} interacting strongly with translational oscillations. At
the transition temperature, this interaction reaches its maximum strength \cite{2} and the effective orientational interaction (rotator-rotator),
which arises from it, determines the nature of the phase transition \cite{10}%
. A detailed investigation of the R-T coupling in solids has been carried
out by Michel. It explains the appearance of the central peak in neutron and
Raman scattering spectra as well as predicts the softening of the transverse
acoustic phonon mode upon approaching the transition temperature, thus
proving the validity of an order-disorder model for the phase transition in
the mixed alkali halide cyanides \cite{11,12,13,13a,13b,14}.

According to Michel's model, the motion of a linear molecular ion in an
octahedral environment can be described by a Hamiltonian that takes into
account the translational and rotational motion as well as their
coupling:\bigskip 
\begin{equation}
H=H^{T}+H^{R}+H^{TR}.  \label{hamilt}
\end{equation}
\noindent The translational and rotational part are the usual ones \cite{11,12,13,13a,13b,14} 
and need not be reproduced here. The term responsible for the
R-T interaction can be approximated as

\begin{equation}
H^{TR}=\sum\limits_{\overrightarrow{k}}i\text{ }Y^{+}\hat{v}s,  \label{TR}
\end{equation}

\noindent where $s(\overrightarrow{k})$ is the Fourier transformed center of
mass displacement of the unit cell and $Y(\overrightarrow{k})$ is an
eigenvector representing rotations. Its components, $Y_{\alpha }$, are
linear combinations of spherical harmonics $Y_{l}^{m}$ of order two: 
\begin{equation}
\vec{Y}=\left( 
\begin{array}{c}
Y_{2}^{0} \\ 
\sqrt{\frac{3}{2}}(Y_{2}^{2}+Y_{2}^{-2}) \\ 
i(Y_{2}^{2}-Y_{2}^{-2}) \\ 
(Y_{2}^{1}-Y_{2}^{-1}) \\ 
-i(Y_{2}^{1}+Y_{2}^{-1})%
\end{array}
\right) .
\end{equation}

\noindent \noindent Two of them have $E_{g}$ and the other three $T_{2g}$
symmetry. The $3\times 5$ matrix $\widehat{v}(\overrightarrow{k})$ reflects
the strength of the coupling and is completely specified in terms of a
microscopic octahedral potential. This R-T coupling leads to an effective
interaction between rotating molecules whose motion is coupled via the
acoustic phonons: 
\begin{equation}
V_{eff}\thicksim \sum\limits_{\overrightarrow{k}}Y^{+}\hat{C}Y,  \label{Veff}
\end{equation}

\noindent where $C(\overrightarrow{k})=v^{T}M^{-1}v$ and $M$ is the coupling
matrix determined by the harmonic part of the translational potential. This
interaction is an addition to the single particle Devonshire potential $%
V^{0} $.

Michel used the above Hamiltonian (\ref{hamilt}) to calculate the elastic
constants $c_{44}$ and $c_{11}$ and found:

\begin{equation}
c_{44}=c_{44}^{0}(1-R_{44}\delta /T)\text{ and }c_{11}=c_{11}^{0}(1-R_{11}%
\gamma /T),  \label{c44c11}
\end{equation}

\noindent where $c_{44}^{0}$ and $c_{11}^{0}$ are the elastic constants in
the absence of coupling; $\delta $ and $\gamma $ are the eigenvalues of the
effective interaction, $V_{eff}$. $R_{44}$ and $R_{11}$\ \ with $T_{2g}$ and 
$E_{g}$ symmetries are defined by a single particle orientational
susceptibility $\hat{R}$: 
\begin{equation}
R_{\alpha \beta }=\frac{Tr(e^{-V_{0}/T}Y_{\alpha }^{+}Y_{\beta })}{%
Tr(e^{-V_{0}/T})}.  \label{Ralpha-beta}
\end{equation}
\ The trace $Tr$ stands for integration over the solid angle. Because $%
\omega _{1}^{2}=\omega _{2}^{2}\varpropto $\ $c_{44}$ and $\omega
_{3}^{2}\varpropto c_{11}$, the frequencies of the corresponding modes can
be written as: 
\begin{equation}
\omega _{1}^{2}=\omega _{2}^{2}=\Omega _{1}^{2}(1-y\delta /T)\text{ and }%
\omega _{3}^{2}=\Omega _{3}^{2}(1-x\gamma /T).  \label{omegas}
\end{equation}

\noindent Here, \noindent $\Omega _{1}$ and $\Omega _{3}$ denote the mode
frequencies in the absence of coupling.

Michel's model has been successfully tested, mainly, on the
alkali-halide-cyanide compounds using various experiments: absorption and
scattering of light \cite{5,6,7,8}, neutron scattering \cite{2,3,4,10}%
, elastic constant measurements \cite{14a}. In all cases, it has been found
to be in good agreement with experimental data. In particular, the elastic
constants have been found to decrease with temperature down to the
transition, in agreement with equation \ref{c44c11}.

\bigskip

As we said, in KTN, a softening of the TA phonon \cite{30,prater} as well as
a decrease in the elastic constants $c_{11}$ and $c_{44}$ \cite{27} has also
been observed. In pure or simple ferroelectrics, a softening of the TA mode
(and corresponding elastic constant $c_{44}$) has been explained in terms of
an interaction between the TA\ and the soft TO ferroelectric mode \cite%
{axe,prater}. However, as our data indicate, in mixed ferroelectrics, the
formation of polar nanoregions with orientational degrees of freedom and the
local distortions and strain fields associated with them, can alter the role
of the classical soft-mode mechanism in the transition. We already mentioned
the unexpectedness of softening of the elastic constant $c_{11}$ for
classical TO-TA mode coupling model. Presented here results show
difficulties of this approach explaining softening of the TA phonon,
especially, below the first transition temperature. On the other hand, we
demonstrate that these contradictions might be resolved by taking into
consideration the orientational dynamics of the polar regions and its
coupling to the TA phonon modes. To make this explanation, we have to
utilize the approach developed by Michel for the disordered crystals, which
do possess orientational dynamics, do not possess a soft optic mode, yet
still exhibit softening of their transverse acoustic mode.

\section{Experiment and Results}

We have studied the temperature dependence of polarized $<$y$|$zz$|$x$>$
(VV) and depolarized $<$y$|$zy$|$x$>$ (VH) Raman scattering spectra of
several $<$100$>$-cut KTa$_{1-x}$Nb$_{x}$O$_{3}$ samples. However, in this
paper, we concentrate primarily on the most characteristic results, which
were obtained on a sample containing 15\% of Nb. The scattering was excited
by 514.5 nm light from a 200 mW Ar$^{+}$-ion laser, focused to a 0.1 mm
spot. The scattered light was collected at an angle of 90$^{\circ }$ with
respect to the incident beam by a double-grating ISA Jobin Yvon spectrometer
equipped with a Hamamatsu photomultiplier R-649. For most of the
measurements, the slits were opened to 1.7 cm$^{-1}$. However, in order to
acquire more precise data in the central peak region, the slits were
narrowed to 0.5 cm$^{-1}$. Both polarizations of the scattered light, VV and
VH, were measured separately. In order to exclude differences in sensitivity
of the monochromator to different polarizations of the light, a circular
polarizer was used in front of the entrance slit. To protect the
photomultiplier from the strong Rayleigh scattering, the spectral region
from -4 to +4 cm$^{-1}$ was excluded from the scans. The data were collected
in several cooling-warming cycles. In each cycle, the sample was cooled from
room temperature down to 80~K and then warmed up, at an average rate of
0.2-0.4~K/min, in temperature steps of 5-10~K. At each step, the temperature
was stabilized and the Raman spectra measured. Cooling and warming cycles
were performed without application of electric field as well as with a 1.2
kV/cm electric field applied in the $z$ direction, perpendicular to the
scattering plane. Below 150~K, the Raman spectra revealed thermal hysteresis
effects. For the sake of clarity, we concentrate on data measured upon
cooling, unless specified otherwise. Both, neutron scattering \cite{28} and
our data indicated phase transitions at the following temperatures: T$%
_{c1}\approx $135 K, T$_{c2}\approx $125 K and T$_{c3}\approx $110 K.
Starting at approximately T$_{c1}$, the sample becomes milky. Near T$_{c3}$
the milkiness suddenly grows to such a degree that all polarization
information is lost.

Raman spectra with polarization analysis of the scattered beam were measured
as a function of temperature. Fig.1 shows examples of the VV (a) and VH (b)
spectra at four temperatures: 200, 140, 130 and 120 K. Due to its low
intensity, the VH component has been magnified by a factor of eight. Upon
lowering the temperature, the first order lines (TO$_{1}$, TO$_{2}$, TO$_{3}$%
, TO$_{4}$) appear on top of a strong and broad second order spectrum of
which the 2TA\ is the most intense feature. For details in line assignment,
see for example \cite{22c,22d,lines,manlief}. Here, we focus our attention
primarily on the second order spectrum and its evolution with temperature.
With decreasing temperature, and transitions to phases of lower symmetries
that restrict reorientational freedom of Nb ions, the VH component of the
second-order spectrum can be seen to decrease dramatically, while its VV
counterpart remains relatively unchanged. The reported decrease is first
noticeable at 140 K and continues until the VH component of the second order
scattering is hardly visible, below 120 K. Therefore, this intensity
decrease correlates with the appearance of the polar phase. Unfortunately,
we were unable to track this effect below the third phase transition
temperature, T$_{c3}$, because the polarization information was lost due to
increasing milkiness of the sample.

Concurrently to the vanishing intensity of the VH component of the 2TA peak,
its frequency can also be seen, in Fig.2, to decrease progressively and
reach a minimum at T$_{c3}$. It is worth noting that the curve is relatively
smooth, and that the two higher transitions produce only slight inflections
in the curve, if at all.

Changes in the rotational freedom of Nb ions are also reflected in the
central peak. Though the central part of the spectrum from -4 to +4 cm$^{-1}$
was cut off, the central peak was found to exceed this range sufficiently so
that it could be fitted adequately and its magnitude and width determined.
The fitting function used was Lorentzian. Fig.3 shows plots of the amplitude
(a) and FWHM (b) obtained from the fits of both VV (solid triangles) and VH
(open triangles) components. An insert illustrates the quality of our
Lorentzian approximation. As seen from the figure, in the high-temperature
limit (cubic phase), the central peak is dominated by the VV component. With
lowering temperature, the growth of the dynamical precursor polar
clusters, at approximately $T\thickapprox 155$~K (T$_{c1}$+20~K), is marked
by a decrease of the VV intensity and a broadening of the central peak in
both VV and VH geometries. With further development of the polar clusters,
the central peak intensity grows and its width decreases. On approach to T$%
_{c3}$, the intensity reaches a maximum and the width, a minimum. The
intensity maximum in VH is even higher than in VV geometry. After the
transition to the R phase, Nb ions become \textquotedblleft locked~in\textquotedblright\ (or confined to) only one site and the central peak
disappears. Comparing with Fig.2, one can see that the cessation of the
reorientational motion changes temperature behavior of the frequency of the
TA phonon.

To get a deeper insight in the processes, we repeated the same set of
measurements in the presence of the electric bias field. Fig.4 presents a
comparative plot where we show examples of data taken at temperatures
150 K (a) and 130 K (b) without field and with a 1.2 kV/cm field applied in
a direction perpendicular to the scattering plane. For a more complete
comparison, we show not only the data taken upon cooling but also the data
taken upon warming the sample up from 80 K. This plot reveals another effect
that we are the first to observe. In both cooling and warming cycles,
application of an electric field causes an increase in the depolarized and a
decrease in the polarized intensity of the second order spectra (lines 1 and
3 to be compared to lines 2 and 4 respectively). The effect of the field is stronger upon
warming, i.e. after the sample has been cooled to 80 K, well into the more
ordered rhombohedral phase.

In the following section, we present an interpretation that consistently
explains all the phenomena described above :

(i) the intensity decrease of the depolarized second order scattering after
transition to the low-symmetry phases upon zero-field cooling;

(ii) the softening of the zone boundary TA mode (seen in Raman spectrum as
2TA peak) on approach to the phase transition temperature;

(iii) the central peak evolution in the temperature region of phase
transitions;

(iv) the field-induced intensity changes in the scattered light.

\noindent

\section{Analysis and Discussion}

The first experimental observation is that the decrease in intensity of the
second-order depolarized spectrum, visible in Fig.1, closely correlates with
the formation and development of polar clusters. The latter formation is
evidenced by the appearance and growth in the spectrum of first order
scattering peaks due to the polar optic modes, TO$_{2}$, TO$_{3}$, TO$_{4}$
and TO$_{1}$. These modes, forbidden by symmetry in the cubic phase, necessarily
signal symmetry breaking at the local level, in this case a tetragonal
distortion \cite{16,17,18,19,20}. With the formation of polar clusters,
new collective degrees of freedom appear, which correspond to the six
possible orientations of the tetragon. Motion between these various
orientations affects the off-diagonal components of the polarizability
tensor and, therefore, influences the VH scattering. The progressive
decrease in intensity of the depolarized components of second order peaks is
consistent with increasing restrictions imposed on the reorientations of the
ions within polar clusters by the successive phase transitions: from 4 in
the T phase to 2 in the O phase and, finally, one in the R phase. 

The second experimental observation of importance is the softening of the
transverse acoustic mode (Fig.2). This softening is consistent with
the slowing reorientational dynamics, strengthened by the nucleation of the polar clusters. 
These polar clusters, with their local strain fields, can especially effectively interact with the
acoustic mode and contribute to the decrease of its frequency. By symmetry, $T_{g}$ rotations couple mostly to the transverse acoustic mode that
corresponds to the $c_{44}$ elastic constant (see eqs.(5) and (6)). According to Michel's concept the softening of the
acoustic mode is described by equation (\ref{omegas}). For our purposes we
rewrite it as 
\begin{equation*}
\omega =\Omega \sqrt{(1-B/T)},
\end{equation*}

\noindent where $B=y\delta $ and $\Omega $ designates the pure 2TA
frequency, uncoupled to rotations. The parameter $B$ is, in general,
temperature dependent but, for short temperature intervals, this dependence
may be neglected \cite{13,13a}. In that case, it has the meaning of a crossover
temperature at which the rotational frequency falls below the frequency of
the acoustic mode. In KTN, as in (KCN)$_{x}$(KBr)$_{1-x}$ for large $x$,
this crossover is never reached because a phase transition causes structural
changes in the crystal. The result of the fit is shown in Fig.2 with two
solid lines. The first line corresponds to the case of cubic phase. The
second line reflects structural changes that develop in the crystal due to the evolution of the polar
phase. It is important to note that, although the experimental curve does
not display any remarkable changes at the phase transition
temperatures, it does show a \textquotedblleft
reduction\textquotedblright\ in softening of the TA mode with the growth of the polar distortions. This \textquotedblleft
reduction\textquotedblright\ indicates that the distorted lattice has new values of the involved parameters, $B$ and $\Omega $. It is reasonable to suppose that the adjustment of these parameters actually occurs in steps, so at each phase transition they obtain new values. However, the resolution of our experiment is not sufficient to observe these details. 

For comparison and for completeness of the picture, we also made an attempt
to approximate this curve by the traditional TO-TA mode coupling approach,
following Axe's work \cite{axe}. To estimate the temperature dependence of
the soft ferroelectric TO$_{1}$ mode we used an approximation of the actual
data measured on the sample containing 15.7\% of Nb \cite{15jt}. It is
remarkable, that according to these measurements, the TO$_{1}$ mode has its
minimum at a temperature approximately 5 K higher than $T_{c1}$. (Similar
measurements on a 1.2\% KTN crystal also showed such a temperature
difference $\thicksim$5~K \cite{16,1.2geh,Chou}). This fact allowed us to
make an adjustement to our value of $T_{c1}$. Consequently, we wrote TO$_{1}$
frequency as:

%\begin{widetext}
\noindent 
\begin{equation*}
\Omega _{TO_{1}}=\left\{ 
\begin{array}{cc}
\sqrt{24(\pm 5)+0.47(\pm 0.05)(T-140(\pm 5)),}\text{ } & T>140K \\ 
11(\pm 1)-0.038(\pm 0.003)T, & T<140K%
\end{array}%
\right. 
\end{equation*}
%\end{widetext}

\noindent Axe's parameters $F_{11}$, $F_{22}$, $F_{12}$ were treated as
free. The result of the fit and the best-fit values of the parameters are
shown in the insert to Fig.2. The first thing that strickes the eye is the
continuation of the TA mode softening till much lower temperatures then
expected from the classical mode coupling theory. It is important to note,
that while the values of $F_{11}$ and $F_{22}$ do agree within the errorbars
to those, predicted by Axe, the value of the parameter $F_{12}$ (which
corresponds to the cross-term, responsible for coupling) is significantly
higher than the one expected by Axe at the zone boundary. This fact reflects
insufficiency of the TO-TA mode coupling theory to explain the acoustic mode
softening even at temperatures above 140 K. Consequently, taking in account
the effect of the reorientational-translational coupling is necessary for
providing an explanation for the softening of the TA mode and can constitute
a challenging problem for a theoretical physicist.\ 

According to Refs. \cite{11,12,13,13a,13b,14}, slowing down of the rotations
should also lead to the growth of the central peak, which is a common
property of systems with slow relaxations. In the present case, the central
peak can be understood in the frame of the modified eight site model, based
on the one described by Sokoloff \cite{23,23a,23b}. As we explained above,
lowering of the temperature leads to appearance of the polar clusters,
transitions to lower symmetry phases, restrictions on the number of allowed
sites and growing correlations between Nb ions. As a result, at high
temperature (Fig.3), a significant contribution of 180$^{\circ }$
reorientations to the scattering process (which, by symmetry, do not
contribute to VH scattering) causes the dominance of the VV component of the
central peak over the VH one. Restriction of these 180$^{\circ }$
reorientations, caused by the nucleation of the polar clusters, leads to a
decrease in intensity of the VV component and to a corresponding increase in
the intensity of the VH one. Concurrently, their widths increase and reach
maxima on approach to the first transition at T$_{c1}$. Below this
temperature, in the tetragonal phase where the sample is divided into
ferroelectric domains, growing correlations between Nb ions lead to an
increase in relaxation time and to the narrowing and growth of both VV and
VH components of the central peak. This process continues through the second
transition at T$_{c2}$ in the O phase. It is remarkable that the VH
component grows faster and becomes even more intense than the VV. Its
intensity reaches a maximum and width a minimum on approach to the third
transition temperature T$_{c3}\approx $110~K. In the rhombohedral phase, the
Nb ions can no longer reorient and the central peak disappears.

This reorientational dynamics coexists with the ferroelectric soft mode TO$%
_{1}$. The peculiarity of this mode in KTN is that it softens only
partially, going through a relatively high energy minimum close to the first
transition. Consequently, only the motion of individual Nb ions is able to
follow the soft phonon dynamics. At the same time, the reorientational
motion of the polar clusters, since it occurs at much lower frequency, can
influence the transverse acoustic phonons frequencies and give rise to the
central peak. Thus, two types of reorientational relaxations are expected
to coexist : (i) the fast relaxation of individual Nb ions, both within clusters
and outside of them, and (ii) the slow relaxation of the entire cluster, i.e. the cooperative
relaxations of all the Nb ions within the cluster (Fig.5). It looks that,
in each phase, precursor clusters of the lower-symmetry phase are present.
Therefore, in the cubic phase (Fig.5a), the two time scales can appear (i)
from the reorientation of Nb ions among equivalent orientations relative to
the polar axis of the precursor cluster and (ii) from the reorientation of
the precursor cluster as a whole relative to the crystal axes. Once the
rotation of the cluster is \textquotedblleft locked\textquotedblright\ in
the T phase (Fig.5b), a precursor distortion leading to the next,
orthorhombic, phase appears. In other words, among four allowed sites, two
become preferred, thus forming a cluster with monoclinic distortion and two
possible orientations. Again, (i) the fast relaxation originates from the
motion between equivalent sites within the same direction of monoclinic
distortion and (ii) the slow relaxation appears due to the reorientational
motion between the two possible monoclinic axes (between different pairs of
equivalent sites).

The observed effect of a bias electric field (Fig.4) confirms the above
interpretation. A field introduces preferred direction in space, along which
the Nb ions and polar regions will tend to orient. In our case, the field
was applied in the $z$ direction, orthogonal to the scattering plane $x-y$
(Fig.5b). As a result, Nb ions became constrained to move only in this
plane, e.g. amongst four equivalent positions. Such a motion, in the
tetragonal and orthorhombic phases, enhances the depolarized scattering
while preventing the polarized one. This effect is clearly seen in Fig.4,
where one should compare the lines 1 and 2 (upon cooling) or the lines 3
and 4 (upon warming), noticing that the effect of the field is
further enhanced by thermal hysteresis.

One of the consequencies of our model is the following: a depolarized
component of second order Raman scattering exists, primarily, due to the
coupling of transverse acoustic modes to the reorientational motion of Nb
ions within polar clusters. To verify this statement, we have measured
samples with low (1.2\%) and high (40\%) concentrations of Nb. In the first
case, at sufficiently high temperature, each Nb ion can be considered as an
approximately free rotator undergoing fast relaxation. It can be shown \cite%
{26}, that the depolarization ratio $I_{VH}/I_{VV}$ ($I_{VH}$ and $I_{VV}$
are the intensities of the depolarized and polarized light scattering) for a
system of free rotators in isotropic space is equal to 0.75. Consequently,
if such a coupling is indeed takes place, we expect to find a depolarization
ratio of 0.75 for the second order peaks too. Lowering the temperature
should result in a decrease of $I_{VH}/I_{VV}$. The experiment did show that
this ratio decreases from 0.75 at room temperature to 0.55 ($\pm $0.05) at
the temperature of transition at $\sim $15~K. In the second case, the
interaction between the neighboring Nb ions is strong even in the
high-temperature phase, so they cannot be considered as free rotators at
all. Therefore, the VH component of scattering will be small and the
depolarization ratio should go to zero. Measurements of the 40\% sample
yielded a value of the ratio of 0.04 (which is on the level of the
experimental error).

We have demonstrated the importance of the reorientational motion of Nb ions as it slows down and the polar clusters form. Various effects, related to their formation, become apparent at different temperatures. Since the relaxor behavior is inseparably related to the presence of the polar clusters, we would like to stop on the question of their development. In the lead relaxors, the temperature of the formation of the polar clusters is often referred to as Burns temperature T$_{d}$ and located several hundred degrees above the temperature of the maximum of the dielectric constant \cite{burns}. One of the effects marking T$_{d}$, is the deviation of the dielectric constant from Curie-Weiss law \cite{viehland-weiss}. In one of the previous works \cite{kpt}, we have shown that in KTN such a deviation occurs approximately at a temperature T$_{c}$+20~K. However, the softening of the TA phonon (Fig.2) and the temperature evolution of the central peak (Fig.3) suggest that the nucleation of the lattice distortions starts at much higher temperatures. A direct confirmation of this suggestion can be found from consideration of the first-order peaks. The most apparent example is given in Ref. \cite{20} (Fig.1a), which shows signs of the presence of a broad reminisce of the TO$_{2}$ peak up to 150 degrees above T$_{c}$. Since in the cubic symmetry this peak is prohibited by selection rules, we have to assume that the distorted regions are present even at this high temperature. The lifetime of these regions must be too short for their detection in the dielectric constant measurements. With lowering temperature, the lifetime of the distortions increases. At a temperature of T$_{c}$+20~K, the polar clusters become sufficiently stable to cause the deviation of the dielectric constant from Curie-Weiss law. The relation between the appearance of the polar clusters and the Burns temperature demonstrates the difference between KTN and lead compounds and establishes a limit of using KTN as a model system for lead relaxors.

Our results indicate that the depolarized second-order scattering is a
sensitive probe for the reorientational motion of both, Nb ions and polar
nanoregions. Analogy to other types of orientationally disordered systems
shows that the rotational-translational coupling is an important and
universal feature of the dynamics of disordered systems. It has been
evidenced and studied in molecular crystals and it may also contribute to an
explanation of the Boson peak observed in the Raman spectra of glassy
materials \cite{24}, which represent a case of complete disorder. Even though
the origin of the Boson peak is still not known, it has been found to be a
universal property of the glassy state and, as we show here, signs of it are
also apparent in disordered crystals. Its origin cannot be explained simply
by acoustic phonons but appears to contain a significant contribution from
localized modes \cite{24}. In glass forming liquids, such as salol, the
depolarized light scattering has been shown to be due to fluctuations in
molecular orientations \cite{25}. We suggest here that there must be a
similar contribution of orientational fluctuations to Raman scattering in
disordered crystals and provide experimental evidence in support of this
claim in the case of KTa$_{1-x}$Nb$_{x}$O$_{3}$.

\section{Conclusion}

The results presented here show the essential role played by orientational
motion in disordered ferroelectric KTN crystals. As in cyanide compounds, it
couples strongly to the translational oscillations and also gives rise to
the central peak. This conclusion, together with earlier results obtained on
glassy materials, shows the fundamental nature of such a coupling and
suggests in return that the Boson peak in glasses may also be due to
coupling between rotational and translational oscillations.

\section{Acknowledgment}

We are grateful to L.A.Boatner for the KTN crystals and to G.Yong for
helpful discussions. This work was supported by a grant from NSF No.
DE-FG02-00ER45842.

\newpage
%\section*{Figure Captions}
\begin{figure}[ht]
\includegraphics[width=1.0\columnwidth]{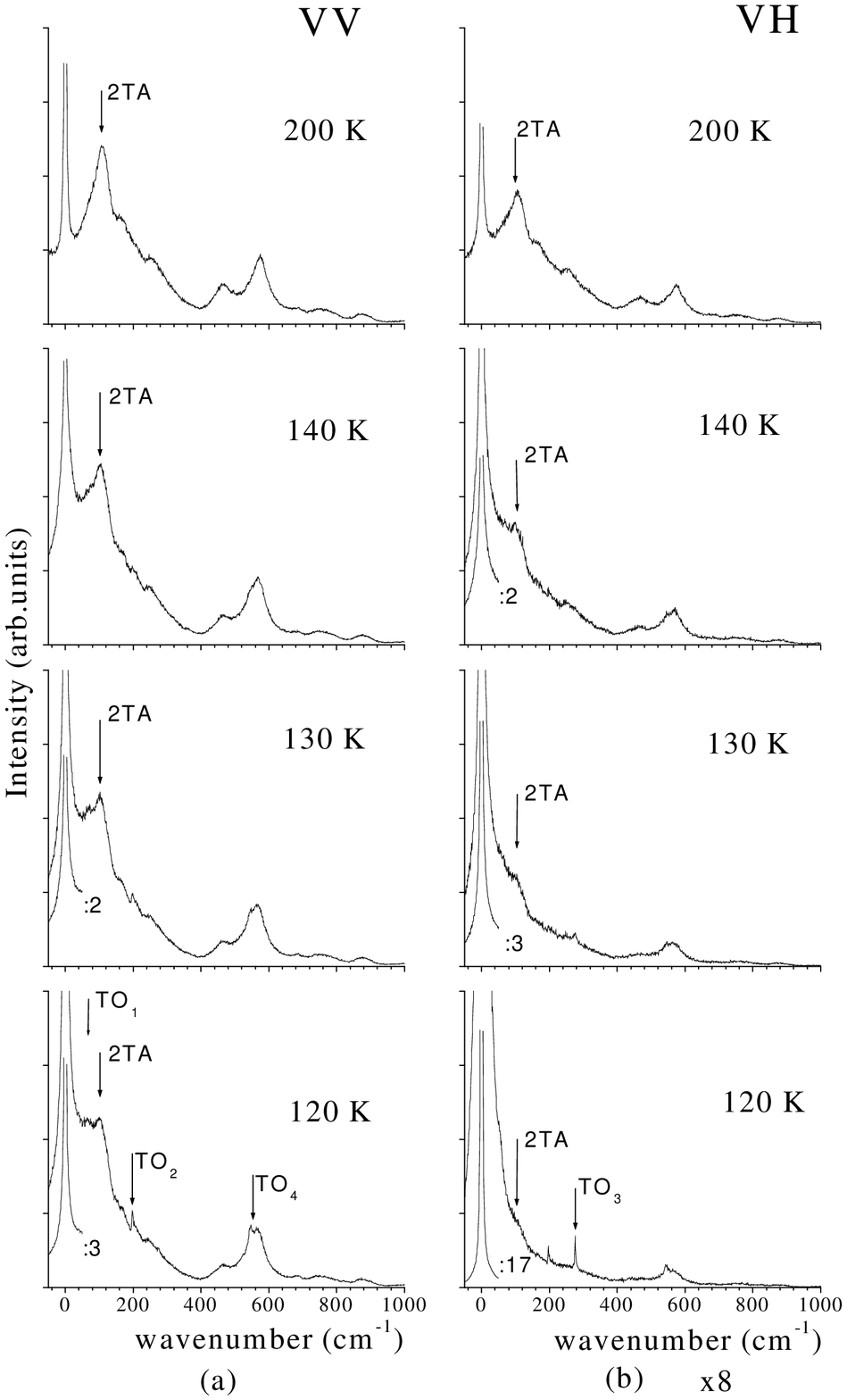}
\caption{VV (a) and VH (b) components of the Raman spectra at different
temperatures.}
\end{figure}
\bigskip

\begin{figure}[ht]
\includegraphics[width=1.0\columnwidth]{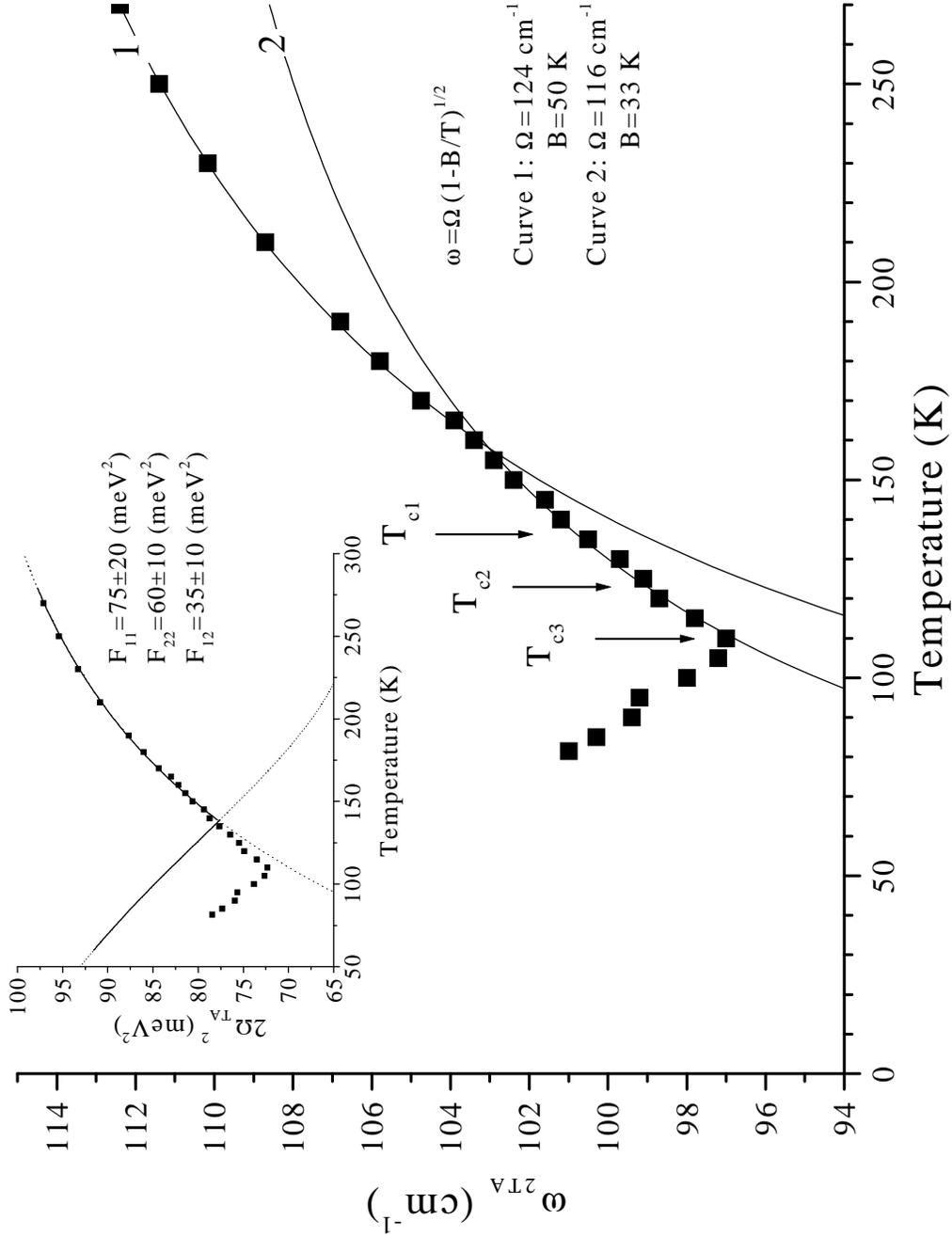}
\caption{Softening of the 2TA mode on the approach to the temperature of the
third phase transition. Squares show experimental data, solid lines -
results of fit based on rotational-translational coupling theory. Appearance
of the polar clusters changes the character of coupling and modifies the
parameters of the fitting curve. Insert shows the best-fit results based on
traditional TO-TA mode coupling theory (values of parameters are outside the
expected range).}
\end{figure}
\bigskip

\begin{figure}[ht]
\includegraphics[width=1.0\columnwidth]{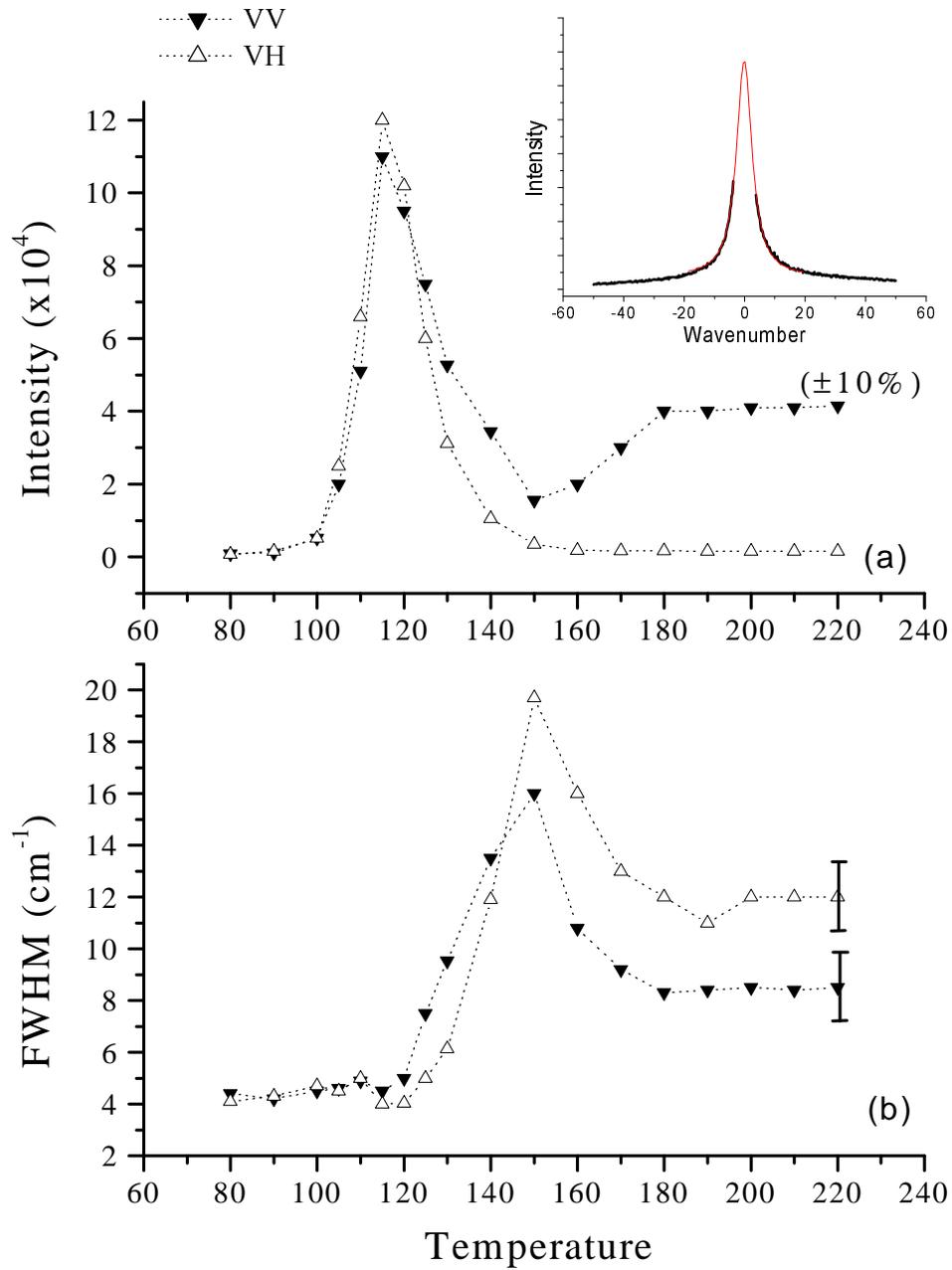}
\caption{Temperature dependence of the intensity and width of the central
peak, approximated by Lorentzian function (example of approximation is shown
on insert).}
\end{figure}
\bigskip

\begin{figure}[ht]
\includegraphics[width=1.0\columnwidth]{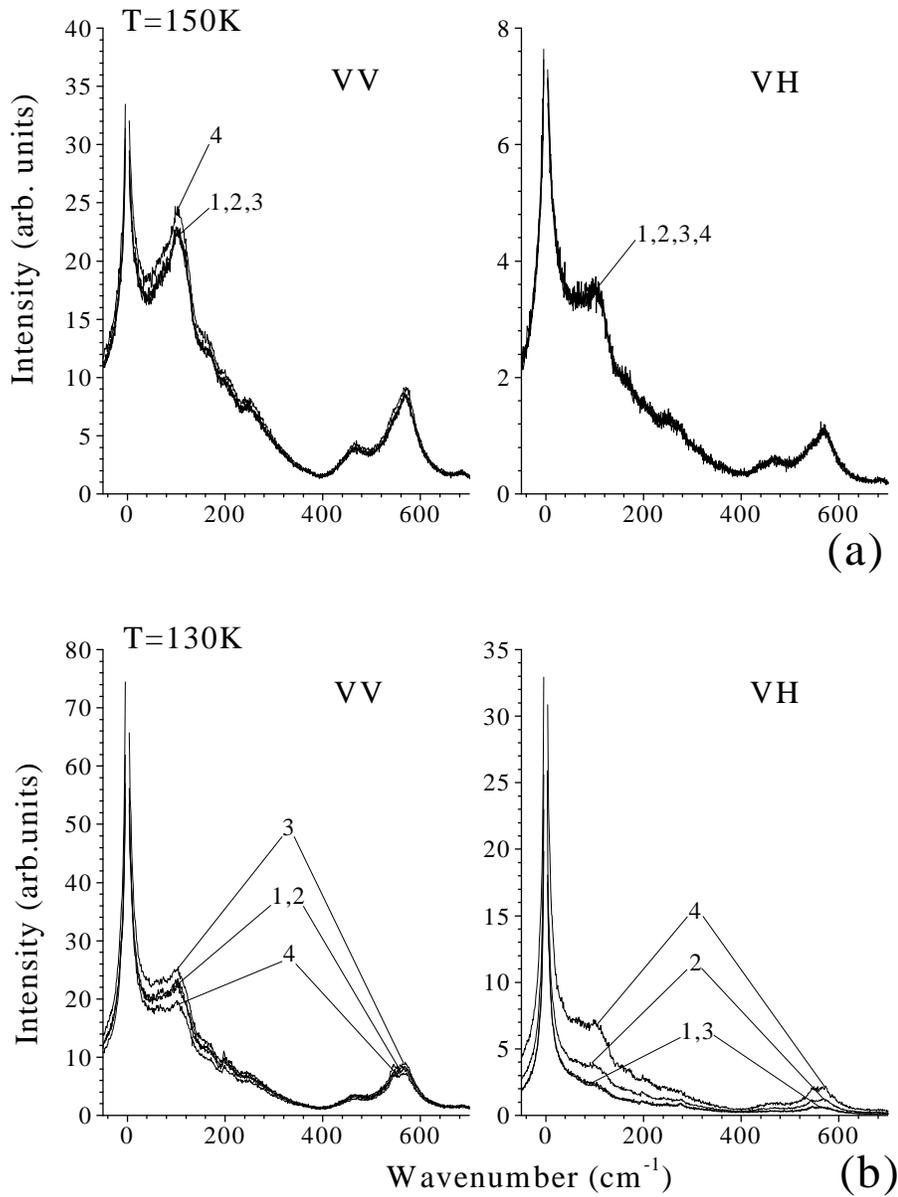}
\caption{An illustration of the influence of electric field measured both, 
upon cooling and warming, on the polarized and depolarized Raman 
scattering at 150 K (a) and 130 K (b). Line 1 is 
measured upon zero field cooling, line 2 upon cooling with 1.2 kV/cm 
electric field. Line 3 is measured upon zero field warming, line 4 upon 
warming with 1.2 kV/cm electric field.}
\end{figure}
\bigskip

\begin{figure}[ht]
\includegraphics[width=1.0\columnwidth]{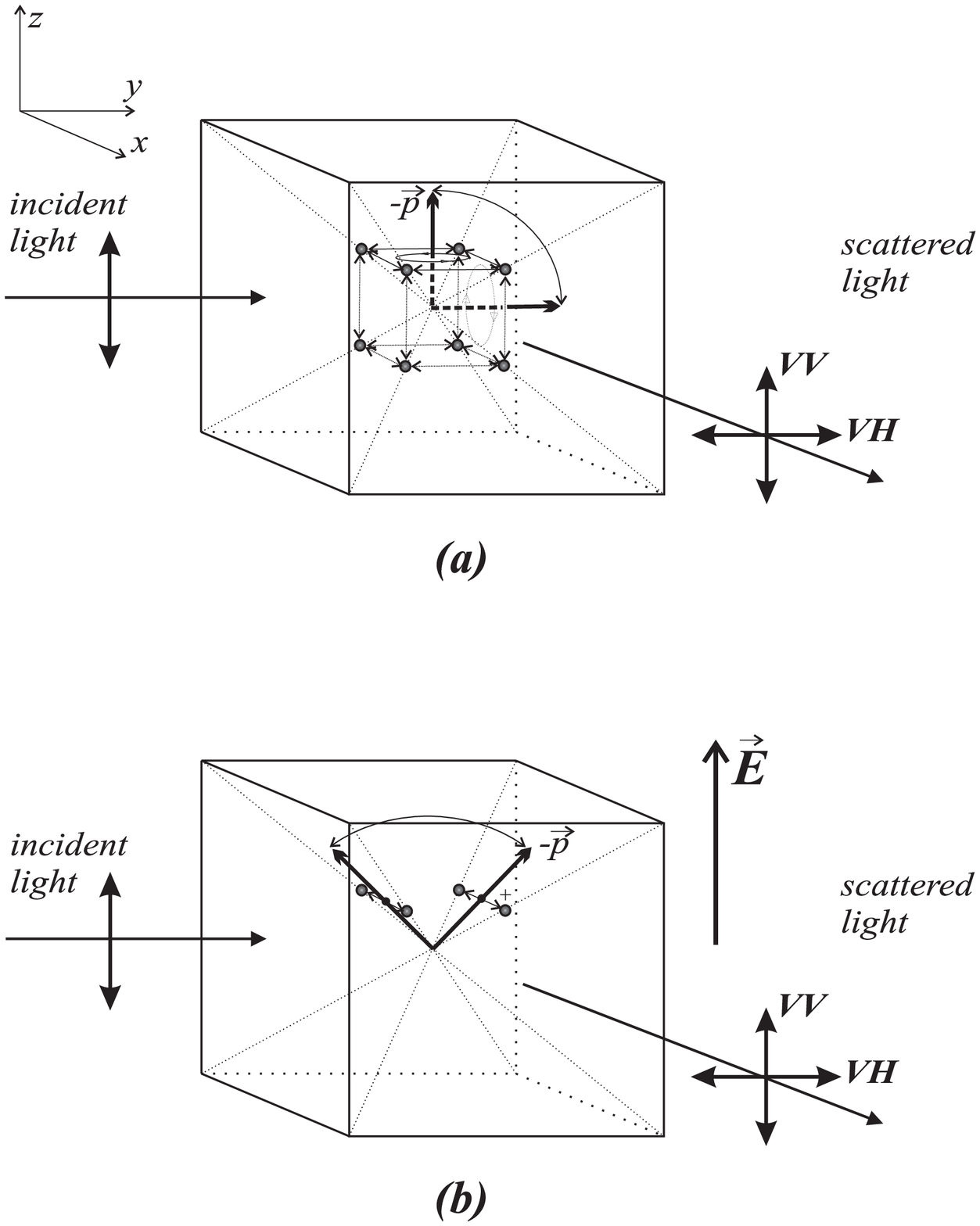}
\caption{Two relaxation time scales: (a) in a precursor polar cluster in the
cubic phase there is a fast motion in (100) plain and slow - among different
planes; (b) in the cluster, "locked" by an external field, a monoclinic
distorsion becomes important; motion is fast along [100] side and slow -
among different sides.}
\end{figure}


\begin{thebibliography}{00}
\bibitem{SinyRev}  I.G.Siny, S.G.Lushnikov, R.S.Katiar, V.H.Schmidt,
Ferroelectrics 226 (1999) 191.

\bibitem{lebon}  A.Lebon, M.El Marssi, R.Farhi, H.Dammak, G.Calvarin, J. Appl. Phys., 89 (2001) 3947.

\bibitem{grace}  G.Yong, J.Toulouse, R.Erwin, S.M.Shapiro, B.Hennion,
Phys.Rev. B 62 (2000) 14736.

\bibitem{radha}  J.Toulouse, R.Pattnaik, J. Korean Phys.Soc. 32 (1998) S942.

\bibitem{rada-res}  R.Pattnaik, J.Toulouse, Phys. Rev. Lett. 79 (1997) 4677.

\bibitem{toul-res}  J.Toulouse, R.Pattnaik, Phys. Rev. B 65 (2001) 024107.

\bibitem{15}  O.Hanske-Petitpierre, Y.Yakoby, J.Mustre De Leon, E.A.Stern,
J.J.Rehr, Phys.Rev. B 44 (1991) 6700.

\bibitem{gr11}  J.J. Van der Klink, D.Rytz, F.Borsa, and U.T.H\"{o}chli,
Phys.Rev. B 27 (1983) 89.

\bibitem{gr12}  E.A.Zhurova, V.E.Zavodnik, S.A.Ivanov, P.P.Syrnikov,
V.G.Tsirelson, Z.Naturforsch 48 (1993) 25; Russ.J.Inorg.Chem. 37 (1992) 1240.

\bibitem{16}  J.Toulouse, P.DiAntonio, B.E.Vugmeister, X.M.Wang, L.A.Knauss,
Phys.Rev.Lett. 68 (1992) 232.

\bibitem{17}  J.Toulouse, R.K.Pattnaik, J.Phys.Chem.Solids 57 (1996) 1473.

\bibitem{18}  P.DiAntonio, B.Vugmeister, J.Toulouse, L.A.Boatner, Phys.Rev.
B 47 (1993) 5629.

\bibitem{19}  E.Bouziane, M.D.Fontana, J.Phys.Chem.Solids 57 (1996) 1473.

\bibitem{20}  P.DiAntonio, J.Toulouse, B.E.Vugmeister, S.Pilzer,
Ferroelectrics Letters 17 (1994) 115.

\bibitem{29}  D.Rytz, H.J.Scheel, J.Cryst.Growth 59 (1982) 468.

\bibitem{21}  P.A.Fleury, J.M.Worlock, Phys.Rev. 174 (1968) 613.

\bibitem{22}  H.Uwe, K.B.Lyons, H.L.Karter, P.A.Fleury, Phys.Rev. B 33 (1986) 6436.

\bibitem{22a}  T.G.Davis, J.Phys.Soc.Jap 28 (1970) 245.

\bibitem{22b}  G.E.Kugel, M.D.Fontana, Ferroelectrics 120 (1991) 89.

\bibitem{22c}  G.E.Kugel, H.Mesli, M.D.Fontana, D.Rytz, Phys.Rev. B 37 (1998) 5619.

\bibitem{22d}  G.E.Kugel, M.D.Fontana, W.Kress, Phys.Rev. B 35 (1987) 813.

\bibitem{22e}  R.Migoni, H.Bilz, D.B\"{a}uerle, Phys.Rev.Lett. 37 (1976) 1155.

\bibitem{lee}  E.Lee, L.L.Chase, L.A.Boatner, Phys.Rev. B 31 (1985) 1438.

\bibitem{lyons}  K.B.Lyons, P.A.Fleury, D.Rytz, Rhys.Rev.Lett. 57 (1986) 2207.

\bibitem{23}  J.P.Sokoloff, L.A.Chase, L.A.Boatner, Phys.Rev. B 41 (1990) 2398.

\bibitem{23a}  J.P.Sokoloff, L.A.Chase, D.Rytz, Phys.Rev. B 38 (1988) 597.

\bibitem{23b}  J.P.Sokoloff, L.A.Chase, D.Rytz, Phys.Rev. B 40 (1989) 788.

\bibitem{axe}  J.D.Axe, J.Harada, G.Shirane, Phys.Rev. B 1 (1970) 1227.

\bibitem{prater}  R.L.Prater, L.L.Chase, L.A.Boatner, Phys.Rev. B 23 (1981) 221.

\bibitem{27}  L.A.Knauss, X.M.Wang, J.Toulouse, Phys.Rev.B 52 (1995) 13261.

\bibitem{Prosandeev}  S.Prosandeev, V.Trepakov, S.Kapphan, M.Savinov, B.Burton, E.Cockayne, 
AIP Proceedings. Fundamental Physics of Ferroelectrics 2002, Williamsburg.

\bibitem{Prosandeev1}  S.A.Prosandeev, V.A.Trepakov, J.Experimental and Theoretical Phys., 94 (2002) 419.

\bibitem{Prosandeev2}  S.A.Prosandeev, V.A.Trepakov, M.E.Savinov, L.Jastrabik, S.E.Kapphan,
J.Phys.: Condens. Matter. 13 (2001) 9749.

\bibitem{1}  K.Knorr, A.Loidl, Phys.Rev. B 31 (1985) 5387.

\bibitem{2}  J.M.Rowe, J.J.Rush, N.J.Chesser, K.H.Michel, J.Naudts,
Phys.Rev.Lett., 40 (1978) 455; J.M.Rowe, J.J.Rush, D.C.Hinks,
S.Susman, Phys.Rev.Lett. 43 (1979) 1158.

\bibitem{3}  J.M.Rowe, J.J.Rush, E.Prince, J.Chem.Phys. 66 (1977) 5147.

\bibitem{4}  J.M.Rowe, J.J.Rush, S.Susman, Phys.Rev. B 28 (1983) 3506.

\bibitem{5}  F.Luty, in Defects in Insulating Crystals, edited by
V.M.Tuchkevich and K.K.Shvarts. Zinatne Publishing House, Riga, 1981, p.70.

\bibitem{6}  F.Luty, Phys.Rev.B 10 (1974) 3677.

\bibitem{7}  D.Durand, F.Luty, Phys.stat.sol. (b) 81 (1977) 443.

\bibitem{8}  D.Durand, F.Luty, Ferroelectrics 16 (1977) 205.

\bibitem{9}  H.U.Beyeler, Phys.Rev.B 10 (1974) 2614.

\bibitem{10}  A.Loidl et al., Z.Phys. B 38 (1980) 253.

\bibitem{11}  K.H.Michel, Z.Phys. B 61 (1985) 45.

\bibitem{12}  K.H.Michel, J.Naudts, B.De Raedt, Phys.Rev. B 18 (1978) 648.

\bibitem{13}  K.H.Michel, J.Naudts, J.Chem.Phys. 67 (1977) 547.

\bibitem{13a}  K.H.Michel, J.Naudts, Phys.Rev.Lett. 39 (1977) 212.

\bibitem{13b}  K.H.Michel, J.Naudts, j.Chem.Phys. 68 (1978) 216.

\bibitem{14}  K.H.Michel, J.Chem.Phys. 84 (1986) 3451.

\bibitem{14a}  J.F.Berret, A.Farkadi, M.Boissier, J.Pelous, Phys.Rev. B 39 (1989) 13451.


\bibitem{30}  P.M.Gehring et al., Ferroelectrics 150 (1993) 47.

\bibitem{28}  G.Yong, J.Toulouse Will be published.

\bibitem{lines}  W.G.Nielsen, J.G.Skinner, J.Chem.Phys 47 (1967) 1413.

\bibitem{manlief}  S.K.Manlief, H.Y.Fan, Phys.Rev. B 5 (1972) 4046.

\bibitem{15jt}  J.Toulouse, R.Pattnaik, Ferroelectrics 199 (1997) 287.

\bibitem{1.2geh}  P.M.Gehring, H.Chou, S.M.Shapiro, J.A.Hriljac, D.H.Chen,
J.Toulouse, D.Rytz, L.A.Boatner, Phys.Rev. B 46 (1992) 5116.

\bibitem{Chou}  H.Chou, S.M.Shapiro, K.B.Lyons, J.Kjems, D.Rytz, Phys.Rev. B 
41 (1990) 7231.

\bibitem{26}  D.A.Long, Raman Spectroscopy. McGraw-Hill Int, New York,
1977.

\bibitem{burns} G.Burns, F.H.Dacol, Solid St. Commun. 48 (1983) 853 and Phys. Rev. B. 28 (1983) 2527.

\bibitem{viehland-weiss} D.Viehland, S.Jang, L.E.Cross, M.Wuttig, Phys. Rev. B 46 (1992) 8003.

\bibitem{kpt} L.A.Knauss, R.Pattnaik, J.Toulouse, Phys. Rev. B 55 (1997) 3472.

\bibitem{24}  C.McIntosh, J.Toulouse, P.Tick, J.Non-Cryst.Solids 222 (1997) 335.

\bibitem{25}  H.Z.Cummins, Gen Li, Weimin Du, R.M.Pick, C.Dreyfus, Phys.Rev.
E 53 (1996) 896.

\end{thebibliography}
\end{document}